\documentclass[aps,pra,twocolumn,showpacs]{revtex4}
\usepackage{graphicx}
\usepackage{siunitx}
\usepackage[pdftex,colorlinks=true,linkcolor=blue,citecolor=blue,urlcolor=blue]{hyperref}
\newcommand{\intensity}[2]{$#1\times10^{#2}\mbox{ W/cm}^{2}$}

\begin{document}

\title{High-Harmonic Generation in the Water Window from mid-IR Laser Sources}

\author{Keegan Finger$^1$, David Atri-Schuller$^1$, Nicolas Douguet$^2$, Klaus Bartschat$^1$ and Kathryn R Hamilton$^1$}
\email{kathryn.r.hamilton@ucdenver.edu}

\address{$^1$Department of Physics  and Astronomy, Drake University, Des Moines, Iowa 50311, USA}

\address{$^2$Department of Physics, Kennesaw State University, Kennesaw, Georgia 30144, USA}

\vspace{10pt}

\begin{abstract}
We investigate the harmonic response of neon atoms to mid-IR laser fields ($2000-3000$~nm) using  
a single-active electron (SAE) model and the fully {\it ab initio} all-electron
\hbox{$R$-Matrix} with Time-dependence (RMT) method. 
The laser peak intensity and wavelength are varied to find suitable 
parameters for high-harmonic imaging in the water window. 
Comparison of the SAE and RMT results shows qualitative agreement between them as well as parameters 
such as the cutoff frequency predicted by the classical three-step model. 
However, there are significant differences
in the details, particularly in the predicted conversion efficiency.  These details indicate the 
possible importance of multi-electron effects, as well as a strong sensitivity of quantitative
predictions on specific aspects of the numerical model.
\end{abstract}

\maketitle

\section{\label{sec:intro}Introduction}

The ``Water Window'' is a region of the electro\-magnetic (EM) spectrum 
spanning from the $K$-absorption edge of carbon at 282~eV 
to the $K$-edge of oxygen at 533~eV. Importantly, water is 
nearly transparent to soft X-rays in this region, while carbon, 
and thus organic molecules, are absorbing. This enables the imaging 
of biologically important molecules in their typical aqueous environments~\cite{biomolecular_imaging}.

Light sources in this region of the EM spectrum are generally produced by 
synchrotron radiation from an accelerator such as a free-electron laser~\cite{water_window_xfel}. 
However, a major problem with this approach is the inability to conduct 
time-dependent imaging of live cells without freezing due to the 
typically low instantaneous power of synchrotron radiation~\cite{xfel_freezing}. 
An alternative approach is high-order harmonic generation (HHG), 
which employs a few-cycle laser system to generate coherent soft X-rays 
by producing odd harmonics of a mid-range infrared (IR) fundamental 
frequency \hbox{\cite{hhg_mid_ir_2014, hhg_mid_ir_2016, hhg_mid_ir_2018}}. 
Such HHG sources could lead to the generation of single-shot absorption spectra 
and live-cell imaging with a femtosecond time resolution~\cite{high_efficiency_water_window_hhg}.

HHG can {\it qualitatively\/} be understood in terms of 
the semi-classical ``three-step model'' suggested by Corkum~\cite{corkum1993}. 
In this model, an electron 1) tunnel-ionizes and 2) is accelerated by a strong electric field. 
As the electric field changes sign, the electron is accelerated back towards its parent ion, where it 3) recollides 
and releases some of its energy in the form of a high-energy photon. The recollision can also excite another electron 
with a higher energy, and the process can be repeated over several cycles of the electric field. 

From the above model, one can obtain a formula for the maximum photon energy, or cutoff energy, 
producible by HHG from a single laser source. This cutoff energy is given by~\cite{corkum1993} 
\begin{equation}\label{cutoff_eq}
    E_{\rm c} = I_p + 3.17\,U_p,
\end{equation}
where $I_p$ is the atomic ionization potential and $U_p$ is the ponderomotive potential. 
The ponderomotive potential can be approximated in terms of the laser peak intensity and wavelength as
\begin{equation}\label{Up}
    U_p = \frac{2 e^2}{c \epsilon_0 m} \cdot \frac{I_0 \cdot \lambda^2}{16 \pi^2}.
\end{equation}
Here $e$ is the electron charge, $\epsilon_0$ the vacuum permittivity, $c$ the speed of light, 
$m$ the electron mass, $I_0$ the laser peak intensity, and $\lambda$ the (central) wavelength of the 
driving laser. 
The pondero\-motive potential scales linearly with the intensity and quadratically with the wavelength.

At first sight, one might just increase the intensity or the wavelength to generate harmonics in the water window. 
However, there are a number of problems with this assumption. Experimentally, increasing the intensity results in 
more ionization, which leads to depletion of the target. On the other hand, if the wavelength is increased instead,
the conversion efficiency decreases, following approximately a $\lambda^{-5}$ or even $\lambda^{-6}$
scaling~\cite{wavelength_efficiency_scaling}. Two or more color fields have been used to extend 
the harmonic cut-off at lower laser wavelengths~\cite{wavelength_scaling_2_color, hhg_synthesised_2_color}, 
but the generation of water-window harmonics typically relies on the use 
of mid-IR lasers~\cite{hhg_mid_ir_2018, 0.5kev_continuua, cep_few_cycle_ir_wwhhg}, 
which suffer from the aforementioned reduced conversion efficiency. However, 
in recent years nanojoule water-window sources have been realized from mid-IR HHG
schemes~\cite{high_efficiency_water_window_hhg,rev_sci_nanojoule_wwhhg}.

The computational side of HHG is not without challenges either. 
The largest of these is an extensive angular-momentum expansion 
needed to obtain partial-wave convergence in a fully quantum-mechanical 
treatment of the process. The computational load for the calculation requires large computers and 
significant run times. Furthermore, longer wavelengths mean larger excursion distances for the electron, 
thus requiring a large configuration space to correctly describe the electron's motion far away from the nucleus.

These aforementioned computational challenges are typically overcome by applying some approximations 
to reduce the overall complexity of the problem. They include the semi-classical strong-field 
approximation (SFA)~\cite{lewenstein} or single-active electron (SAE) models~\cite{sae_hhg_ivanov}, 
both of which neglect multi-electron effects. 
However, such effects have been shown to significantly change the harmonic 
cutoff and/or the intensity of the generated harmonics through phenomena such as the giant 
resonance in xenon~\cite{santra_xenon_4d}.
Multi\-electron effects on the production of HHG were reported in a variety
of studies over the years~\cite{PhysRevLett.96.223902,PhysRevA.87.062511,
PhysRevA.94.023405,PhysRevLett.118.203202,PhysRevA.95.033402}, including
recent works on
He~\cite{Wozniak2022} and Be~\cite{Kutscher2022}.

The $R$-Matrix with Time-dependence (RMT) method accounts for 
multi-electron effects in HHG by employing a highly efficient code 
specifically designed for high-performance computing \cite{RMT_CPC}. 
RMT is indeed suitable for serious computational challenges such as HHG from mid-IR lasers, 
and has previously been used to describe HHG from 1800~nm sources~\cite{ola_nearIR}. 
In the present project, we significantly extend previous RMT calculations by employing driving laser 
wavelengths of up to 3000~nm, resulting in harmonic cutoffs 
within the water window for realistic peak intensities of the driving laser. 

The harmonic spectra from our RMT calculations 
are also compared with those obtained by an SAE approach. This provides 
us with means to determine the influence of multi-electron effects, as well as other aspects
of the numerical model and the subsequent extraction of the results, 
on the predicted spectra in this energy region.  Furthermore, when pushing a complex method like RMT 
into previously unchartered territory, it is certainly advisable to have an entirely different
approach available for comparison.

Section~\ref{sec:methods} gives an overview of the numerical methods and the parameter space investigated.
This is followed by the presentation and discussion of the predicted 
harmonic spectra for a neon atom in laser fields of various intensities and wavelengths in 
Sect.~\ref{sec:results}. Conclusions and an outlook are given in Sect.~\ref{sec:ch5conclusion}.
Unless specified otherwise, atomic units (a.u.) are used below.

\section{\label{sec:methods}Numerical Methods}
Below we describe the methods used in the present study.  
We start with a brief summary of our own SAE approach in
sub\-section~\ref{sec:sae},
which is followed by a more detailed description of the RMT method in~\ref{sec:rmt}.  
The parameter space investigated is presented in sub\-section~\ref{sec:calcparameters}.  The section concludes with
a discussion of ``windowing'' in~\ref{sec:windows}, a technique that has been widely used to improve the 
qualitative appearance of the final results. However, we then show that there can be substantial changes
in the quantitative predictions, such as the conversion efficiency, if such windows are applied.  Consequently,
they should only be employed with great care and most likely not when absolute values of the spectral 
HHG density are of interest.  

\subsection{\label{sec:sae}The Single-Active-Electron Approach}
We use the same basic method and the associated computer code 
as described by Pauly {\it et al.}~\cite{PhysRevA.102.013116}. Briefly, we solve the time-dependent
Schr\"odinger equation (TDSE) for the active electron, which is initially in the $2p$ orbital of the
neon ground state.  We use the potential suggested by Tong and Lin~\cite{Tong_2005}, 
in which we calculate the $2p$ and
other bound orbitals (only used here to check the quality of the structure description), 
as well as the continuum orbitals to represent the ionized electrons.  The above
potential produces the first ionization potential of neon, i.e., the binding energy of the $2p$ electron,
very well.  We employ a finite-difference method with a variable radial grid, with the smallest
stepsize of 0.01 near the nucleus and a largest stepsize of 0.05 for large distances.  The timestep is
held constant at 0.005. The calculations are carried out in the length gauge of the electric dipole operator.
Partial waves up to angular momenta~$\ell = 300$ were included in order to ensure converged results. 
Tests carried out by varying the time step, the radial grid, and the number of partial waves give us 
confidence that any inaccuracies in the predictions are due to the inherent limitations of the SAE model rather
than numerical issues.

We then calculate $z(t) \equiv \langle \Psi(\textbf{r},t)|z|\Psi(\textbf{r},t)\rangle$, i.e., effectively
the dipole moment for linearly polarized light with the electric field 
vector along the $z$~direction, at each time step and numerically 
differentiate it once with respect to time to obtain the dipole velocity $v(t)$
and again to obtain the dipole acceleration~$a(t)$.  Following 
Joachain {\it et al.}~\cite{Joachain} and Telnov {\it et al.}~\cite{PhysRevA.87.053406},
the spectral density~$S(\omega)$ is obtained as
\begin{equation}
S(\omega) \!=\! \frac{2}{3\pi c^3} \,|\tilde{a}(\omega)|^2 \!=\! 
\frac{2}{3\pi c^3}\, \omega^2 \,|\tilde{v}(\omega)|^2 \!=\! \frac{2}{3\pi c^3}\, 
\omega^4 \,|\tilde{z}(\omega)|^4,
\label{joachain}
\end{equation}
where
\begin{equation}\label{fourier}
\tilde{a}(\omega) = \int_{-\infty}^{+\infty} a(t) \,{\rm e}^{i \omega t} dt
\end{equation}
with similar definitions for $\tilde{v}(\omega)$ and $\tilde{z}(\omega)$.  
The factor $1/\sqrt{2\pi}$ factor that makes the Fourier transform symmetric is absorbed by the
pre\-factor in Eq.~(\ref{joachain}), whose numerical value is approximately $8.25 \times 10^{-8}$ 
in atomic units. When a discrete Fourier transform is used, the resulting spectrum needs to be multiplied by the appropriate scaling factor to
account for the time grid on which the sample function is available. 

Next, we found that using $\omega^4 |\tilde{z}(\omega)|^2$ without further manipulation causes numerical
problems, since the distribution of the ejected electrons leads to a 
remaining net dipole moment at the end of the pulse, which actually varies slowly with time. 
If the above method is used, the
acceleration form is least affected and results in the sharpest cut\-off.

Finally, we need to account for the fact that the actual ground state of 
Ne is $(1s^2 2s^2 2p^6)^1S$.  Hence, if we 
look at the $2p$ electron, we need to perform calculations for the possible 
magnetic sublevels, i.e., $m=0$ and 
$m = \pm 1$, respectively.  Due to the symmetry properties of the process, 
the results for $m = -1$ and $m=+1$ for a linearly 
polarized driving field
are identical.  In order to account for all six electrons, we multiply the dipole moment for the sublevels by
2~for $m=0$ and 4~for $m=1$, respectively, which leads to factors of 4 and 16 in the spectral density.  
We will see below, however, that for the process of interest in this manuscript, the
contribution from $m=0$ dominates that from $m=\pm 1$.  Hence, in practice it is sufficient to only perform
the calculation for $m=0$.  

Even though these aspects are not relevant for the {\it qualitative\/}
description of the spectrum, we believe it is important for theory to put the
predictions on an absolute scale using the procedure outlined above.  This also allows to calculate
the conversion efficiency
\begin{equation}\label{efficiency}
C \equiv \int_{-\infty}^\infty S(\omega) \,d \omega \; {\mbox{\Large /}} \int_{-\infty}^\infty E^2(t) \,dt.
\end{equation}
We calculate the latter for both the entire spectrum (where the contribution is dominated by
the first harmonic) and the plateau region, which is the important part in generating
the high harmonics.  

We also note in passing that i)~the first harmonic does not peak at
exactly the same wavelength as the central driving frequency for short pulses and ii)~the
plateau is not really flat for the extreme parameters shown in this paper.  These outcomes are due to 
the non\-vanishing contribution from the time derivative of the envelope function and  
complex interference effects from emission at slightly different times. Since the
induced dipole moment follows the driving field almost adiabatically, which would suggest the dominant
frequency in the emission spectrum to be the central frequency of the driving field, 
the peak is actually shifted to slightly higher frequencies.

Furthermore, the theoretical spectrum is often smoothed to improve its look by convolving it with
something that might resemble the energy resolution of a spectrometer.  We decided to refrain from
doing this and will present our original results below.

\subsection{\label{sec:rmt} The R-Matrix with Time-Dependence Method}
RMT is an {\it ab-initio} technique that solves the time-dependent Schr\"odinger equation by employing 
the \hbox{$R$-matrix} paradigm, which divides the configuration space into two separate regions over the 
radial coordinate of a single ejected electron~\cite{RMT}.

In the inner region, close to the nucleus, the time-dependent $N$-electron wave function is represented
by an $R$-matrix (close-coupling) basis with time-dependent coefficients. In the outer region, 
the wave function is expressed in terms of residual-ion states coupled with the radial wave function 
of the ejected electron on a finite-difference grid. The solutions in the two separate regions are 
then matched directly at their mutual boundary. The wave function is propagated in the length gauge 
of the electric dipole operator. For the typical atomic structure descriptions used in RMT, this 
converges more quickly than the velocity gauge \cite{tdrm_dipole_gauge}.  On the other hand,
propagation in the length gauge usually requires more angular momenta to obtain 
partial-wave-converged
results~\cite{Cormier_1996,PhysRevA.81.043408}. Investigating the optimum procedure for long wavelengths, therefore,
might be a worthwhile future project.

In the RMT calculations presented here, the neon target is described within an $R$-matrix inner 
region of radius 20 and an outer region 
of up to 65,000. The finite-difference grid spacing in the outer region is 
0.08 and the time step for the wave-function propagation is 0.005, 
except for one case. We found that
careful checks of these numerical parameters need to be performed, if quantitative rather than 
qualitative results are desired.

The target was described with the single-electron orbitals generated by 
Burke and Taylor~\cite{window_resonances_argon}, using the \hbox{RMATRX-II} suite of codes~\cite{rmatrix_II}.  In order to assess the sensitivity of our predictions,
we set up close-coupling expansions with just the ionic ground state $(2s^22p^5)^2P$ (labelled ``1st'') or 
a 2-state (2st) model to include the most important coupling to the $(2s2p^6)^2S$ state.  Next, we checked
the dependence of our predictions on the number of configurations employed to describe these states. Specifically,
we tested just the ground state with its dominant configuration (1st-1cf), an improved description involving
seven configurations (1st-7cf), the \hbox{2-state} model of Burke and Taylor (2st-13cf) with 13 configurations, 
and finally a \hbox{2-state} model with 23 configurations (2st-23cf).  
The latter is the best we could do with the available computational resources. 

We then included all available $2s^22p^5\epsilon \ell$ and $2s2p^6\epsilon \ell$ 
channels up to a maximum total orbital angular momentum of $L_{\mathrm{max}}=240$. 
For the majority of the calculations presented here the continuum functions were constructed from a set of 60 \hbox{$B$-splines} of order 13 for each angular momentum of the outgoing electron. For the results presented in Fig.~\ref{fig:3000nm}, corresponding to the highest cutoff energies achieved in this work, 64  splines and a timestep of 0.003 were
needed for convergence.

The harmonic spectra from RMT calculations are obtained by Fourier transforming and squaring the
time-dependent expectation value of the dipole operator~\cite{brown_helium}, and scaling by 
the appropriate power of $\omega$. Both the length and 
velocity forms can be applied to obtain the final spectrum, 
as the length and velocity matrix elements generated from the time-independent structure 
calculation can be utilized by the RMT code. 
We emphasize that the RMT results in the velocity gauge are {\it not} those obtained by 
differentiating the time-dependent dipole moment from the length form with respect to time. Instead, 
the necessary matrix elements are calculated independently.  Consequently, the level of 
agreement between the results obtained in the two gauges provides some (albeit indirect)
assessment regarding the numerical quality of the implementation.  To check our results once again,
we never\-theless performed the differentiations and thereby obtained two versions of the velocity 
results, one directly from the velocity-matrix elements and one by differentiation of 
the length-form result.  Similarly, we generated
two versions in the acceleration gauge.  The overall excellent agreement between these results provides
confidence in their numerical accuracy.

\subsection{\label{sec:calcparameters}Calculation Parameters}

The primary laser pulse is a 6-cycle, 2000-3000~nm linearly polarized 
pulse with peak intensities ranging from \intensity{1.0}{14} to \intensity{2.0}{14}. 
A 3-cycle, sine-squared ramp-on/off profile approximates the more realistic Gaussian 
profile sufficiently well, while providing some numerical advantages over the Gaussian. 
The short duration of the pulse was chosen to isolate the generation of the cutoff 
harmonics to a single trajectory of a single IR cycle~\cite{hamilton_two_colour} 
and to mimic a recent experiment that used HHG from mid-IR sources to generate soft-X-ray pulses~\cite{43as_pulse}.
We used a ``cosine pulse''
with the carrier envelope phase chosen to ensure a vanishing displacement, i.e., 
$\int_0^T A(t)\,dt = 0$.

When trying to simulate an experimental setup even more realistically, the carrier envelope
should be averaged over.  In order to avoid possibly unphysical pulses when doing so, it is
better to set the vector potential instead~\cite{PhysRevA.90.043401} and calculate the
electric field via its time derivative.  The RMT calculations performed for this work, however,
are computationally too expensive to do the averaging with our currently available computational resources.
Hence, we decided on the present procedure and thus avoided a further disturbance of the sinusoidal
electric field that would result from setting the vector potential in a simple form and then
also differentiating the envelope function.

Calculations were performed on the Stampede2~\cite{stampede2} and Frontera~\cite{frontera} 
supercomputers, both based at the Texas Advanced Computing Center at the University of Texas at Austin, 
and Bridges-2~\cite{bridges2} at the Pittsburgh Super\-computing Center. Each RMT calculation 
required about 500~cores for 8-10~hours, with variations depending on intensity and wavelength. 
Calculations performed with the SAE code took at most a few hours on a single node
using all available cores of that node via OpenMP parallelization.

\subsection{\label{sec:windows}Windowing}
Another issue worth mentioning concerns the use of windowing functions. In the RMT suite of codes~\cite{RMT},
a Gaussian mask is applied to the wave function when determining the expectation value of the dipole length, 
ensuring that its calculation is limited to the physically most meaningful region of almost no ionization, 
i.e., excursion ranges from where the electron can actually return.  
Thus we can directly determine the time-varying expectation values of 
both the dipole operator and the dipole velocity. 

\begin{figure}[!b]
    \centering
    \includegraphics[width=0.98\columnwidth]{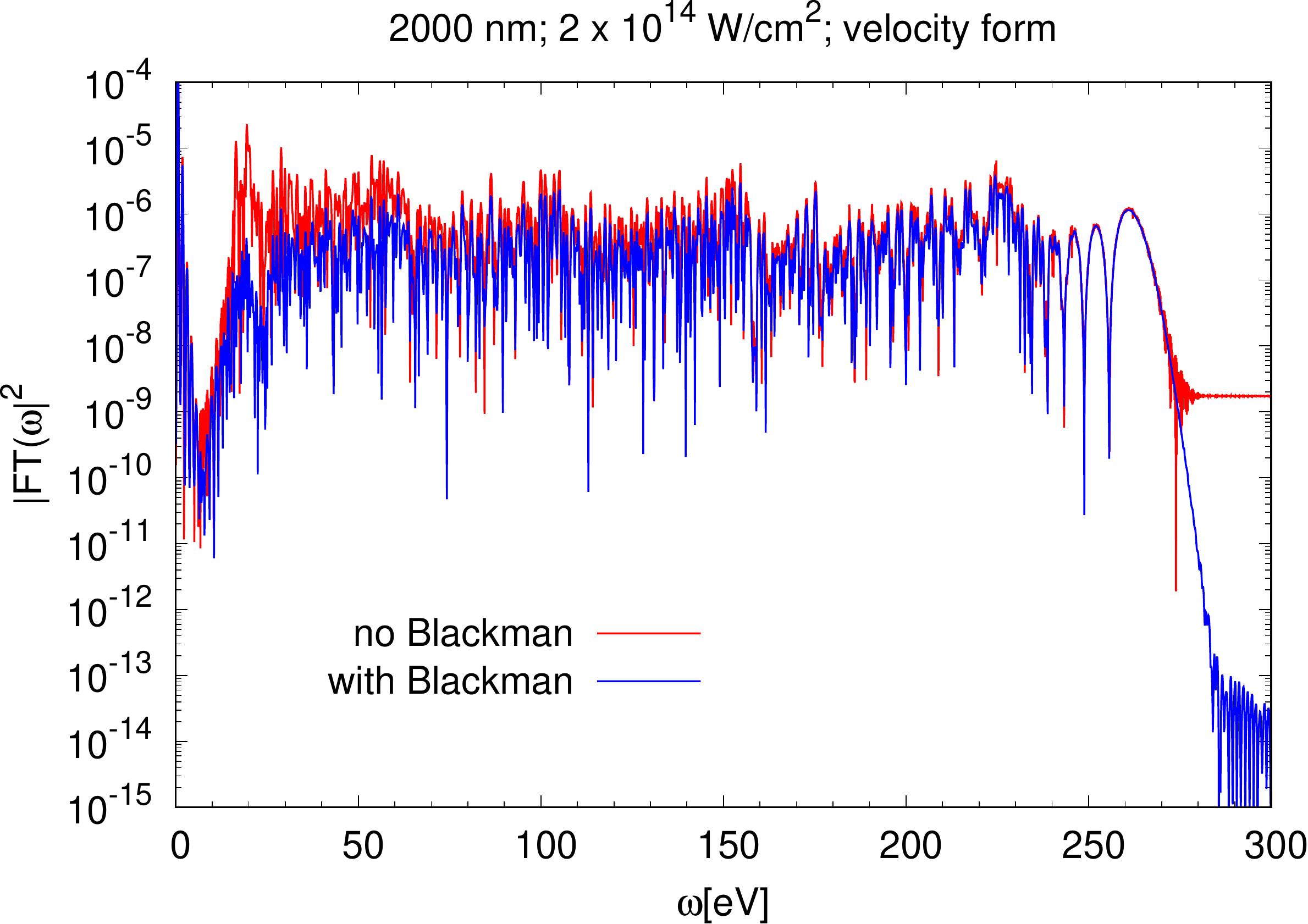}
    \caption{Harmonic spectrum of neon in a 2000~nm, \intensity{2.0}{14} laser pulse, 
    as obtained with RMT in the velocity form of the electric dipole operator with and
    without the Blackman window.  The window generally reduces the result.
    The harmonic plateau extends from about 30~eV to a cutoff of about 260~eV.
    }
\label{fig:Blackman}
\end{figure}

After calculating the dipole expectation values, typically a Blackman window~\cite{blackman} 
is applied to ensure a smooth trend of both the dipole moment 
and its time derivative towards zero at the end of the pulse.
Using the Blackman window significantly reduces the numerical noise arising 
from the Fourier transform needed to calculate the harmonic spectrum, 
and also improves the visibility of the cutoff. 
On the other hand, for short pulses where the dipole moment does not 
exhibit sufficient repetition, the window severely
affects the absolute value of the spectral density, as shown in Fig.~\ref{fig:Blackman}.
The reduction in the absolute values is particularly evident at the beginning of the ``plateau'' from
about $20-50$~eV, where it can exceed an order of magnitude and will severely affect the integral under the curve. 
Since we are interested in obtaining a {\it quantitative\/} 
value for the conversion efficiency, which is practically
independent of both the steepness of the cutoff and the depth of the 
minimum in the spectral density at very low energies, we do not use this window in 
the present work.

Finally, we address the issue of which form is most appropriate to show the
obtained spectra. For the SAE calculations, using the acceleration form 
after differentiating the induced dipole moment twice with respect to time provides the
numerically cleanest results.  The velocity form gives very similar results, except for the 
cut\-off not being as deep and small changes in the depth of the low-energy minimum.  
Using the length form as written in Eq.~(\ref{joachain}), however, is
problematic, as mentioned above.  Indeed, a window like that used in the RMT code to limit the integration
range for the calculation of the dipole moment fixes this problem, but it is not necessary
when using the acceleration (or velocity) forms. 

The RMT predictions obtained
in both the length and velocity form directly give very similar results, as long as a
window is used when the dipole moment is generated.  As discussed by
Baggesen and Madsen~\cite{Baggesen_2011}, from a quantum-mechanical point of view, the velocity
form might be the most natural one.  Hence, most of the RMT calculations presented below
were obtained directly with the matrix elements calculated in that form.  The small price
to pay is the less steep drop in the spectral density at the cutoff frequency, which however
has practically no effect on the integral under the curve taken up to or even slightly above
the cutoff.

As seen from the above discussion, a quantitative calculation of the spectral density, 
which is needed to obtain the conversion efficiency, is significantly more challenging
than a purely qualitative description. All implementations we employed, be it SAE 
with a local potential or RMT
with either one or two coupled channels, with simple or more sophisticated target wavefunctions,
will predict a dominant peak in $S(\omega)$ at almost the driving frequency.  This peak is
followed by a deep minimum after about the fifth harmonic in our cases, a plateau, 
and finally a cut\-off near the energy predicted by the three-step model~\cite{corkum1993}. 

\section{\label{sec:results}Results and Discussion}

The first step in determining the efficacy of RMT in producing water-window harmonics is to ensure 
that the length and velocity forms of the produced spectrum agree reasonably well.

\begin{figure}[!htb]
    \centering
    \includegraphics[width=0.98\columnwidth]{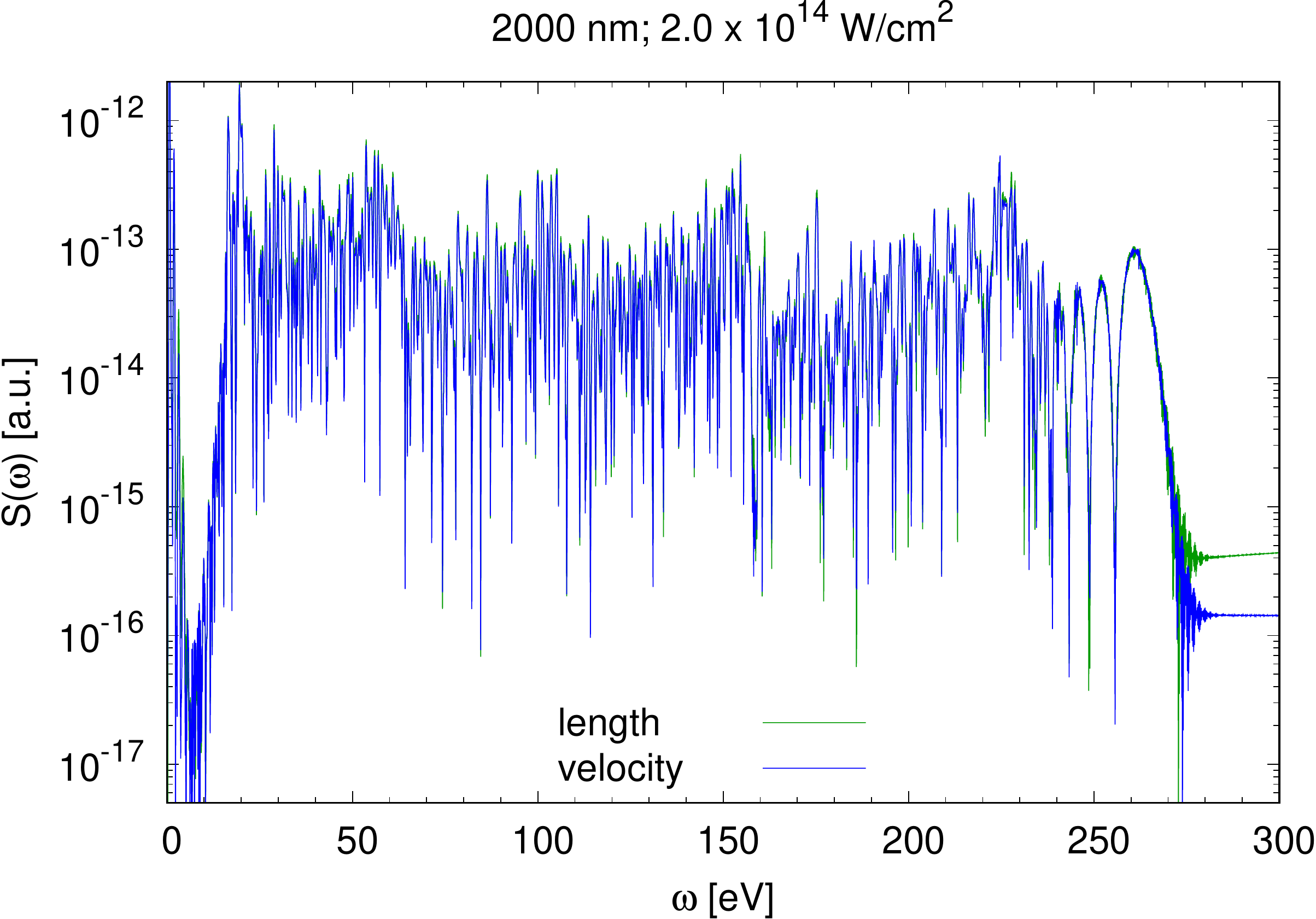} \bigskip
    \includegraphics[width=0.98\columnwidth]{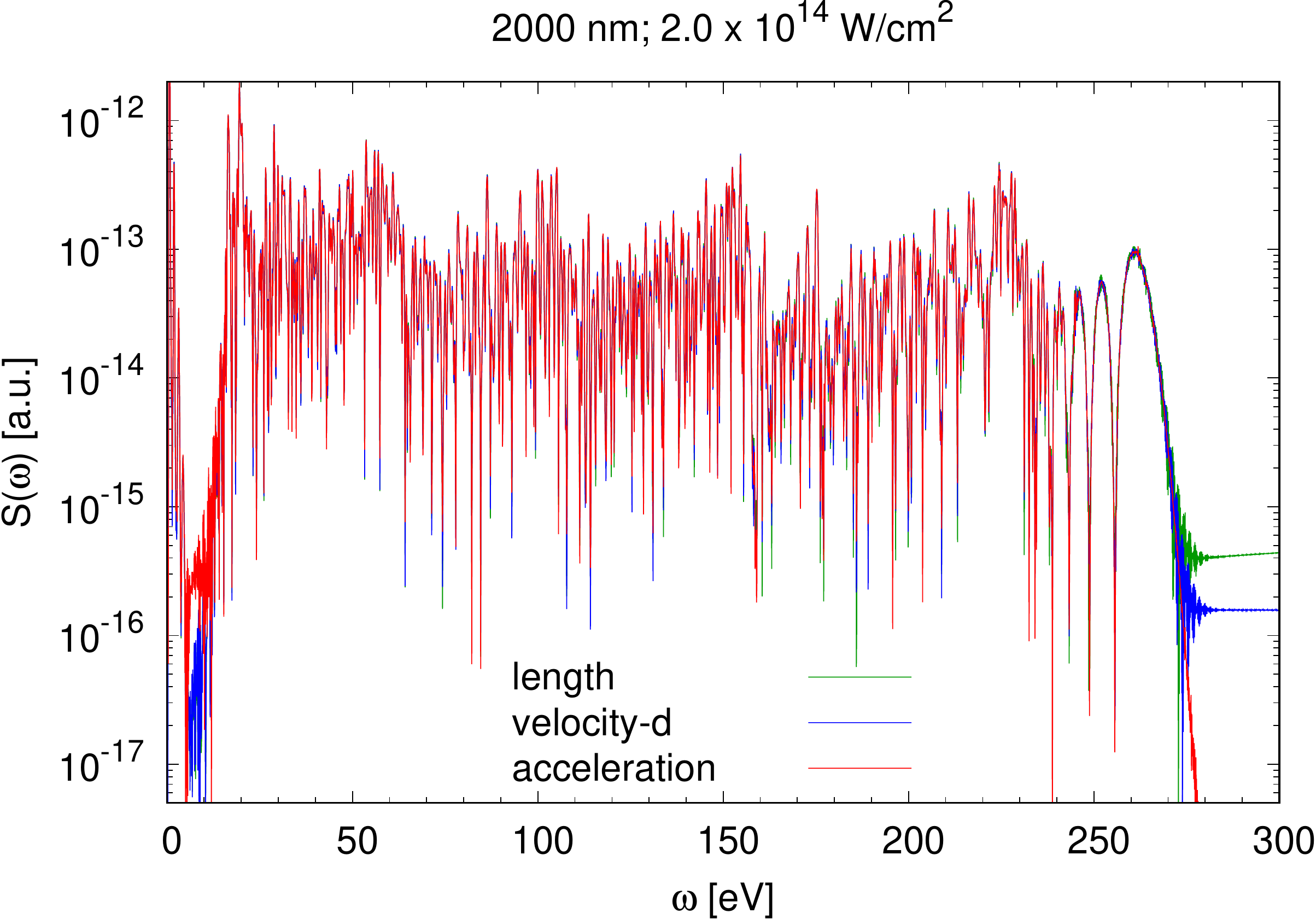}
    \caption{Harmonic spectrum of neon in a 2000~nm, \intensity{2.0}{14} laser pulse.
    The top graph shows the RMT results obtained directly with the length and velocity 
    matrix elements of the dipole operator, with the latter dropping off more after the
    cut-off. The bottom part shows the 
    results obtained from the matrix elements in the length form (highest after the cut-off)
    and then
    the induced dipole moment differentiated (``-d'') once to get the velocity (middle line
    after the cut-off) and 
    acceleration results (deepest after the cut-off).  See text for details.
    }
    \label{fig:gauges}
\end{figure}
Figure~\ref{fig:gauges} depicts a typical spectrum obtained in five different ways. 
Specifically, we compare the {\it independent\/} RMT results obtained with dipole matrix
elements generated by \hbox{RMATRX-II} in the length and velocity gauges (top part of Fig.~\ref{fig:gauges})
and those obtained by differentiation of the dipole moment in the length form (bottom part).
Except for the steepness of the cutoff and the depth of the
low-frequency minimum, all the results are very similar and almost indistinguishable
on the graph.  

\begin{figure}[!htb]
    \centering
    \includegraphics[width=0.98\columnwidth]{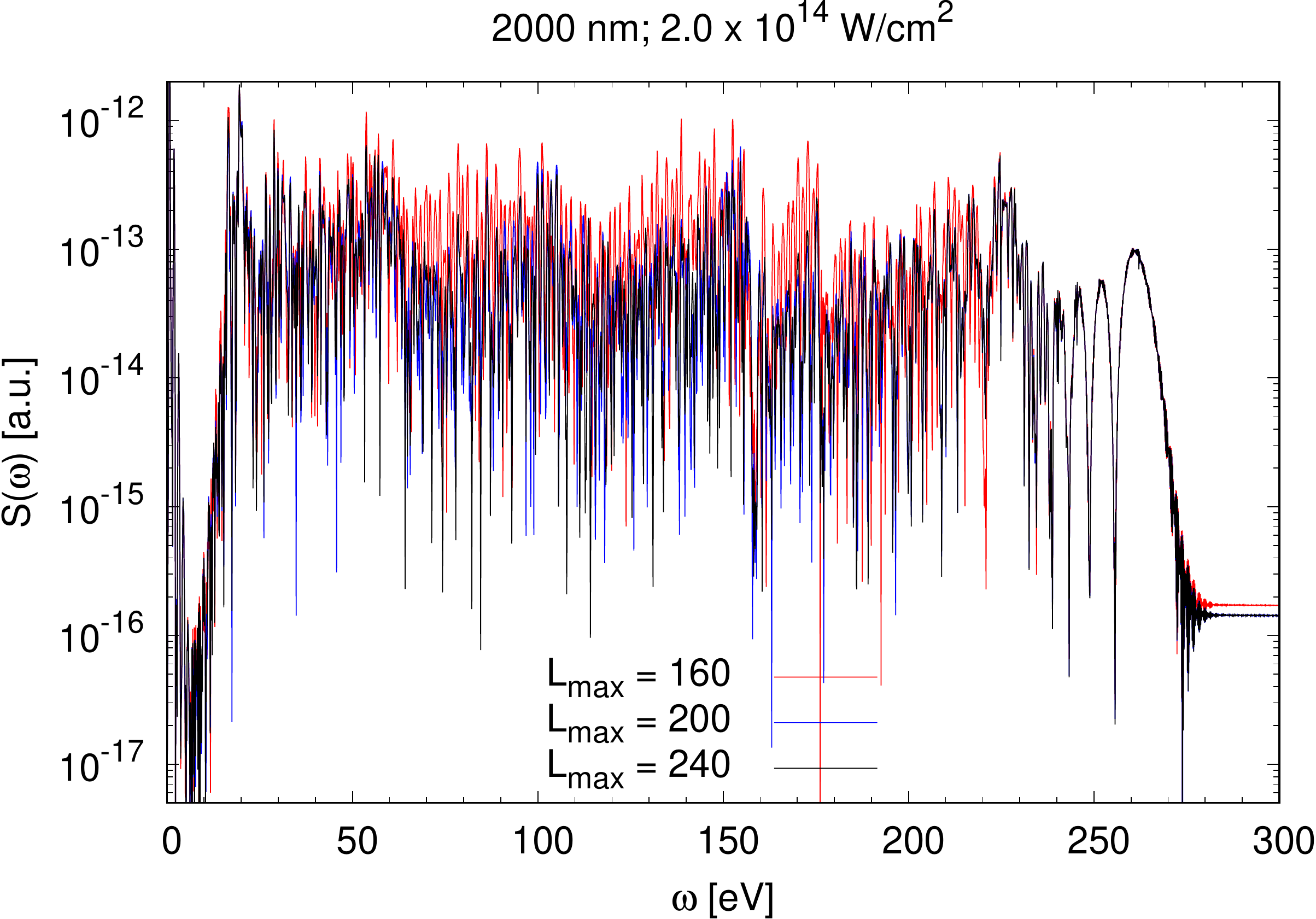}
    \caption{Harmonic spectrum neon in a 2000~nm, \intensity{2.0}{14} laser 
    pulse obtained with different values of $L_{\mathrm{max}}$. 
    }
    \label{fig:lmax_converge}
\end{figure}
Due to the large angular-momentum expansion needed for the calculation of multi-electron effects, 
it is also important to verify that the chosen value of the largest term ($L_{\mathrm{max}}$) does not 
affect the convergence of the results. This test is performed by running several calculations while 
increasing $L_{\mathrm{max}}$. A converged calculation is shown in Fig.~\ref{fig:lmax_converge}, 
where the results obtained with values of $L_{\mathrm{max}}=200$ and $L_{\mathrm{max}}=240$ are very similar, while
differences with those obtained with $L_{\mathrm{max}}=160$ are still visible.

\begin{figure}[!tb]
    \centering
    \includegraphics[width=0.98\columnwidth]{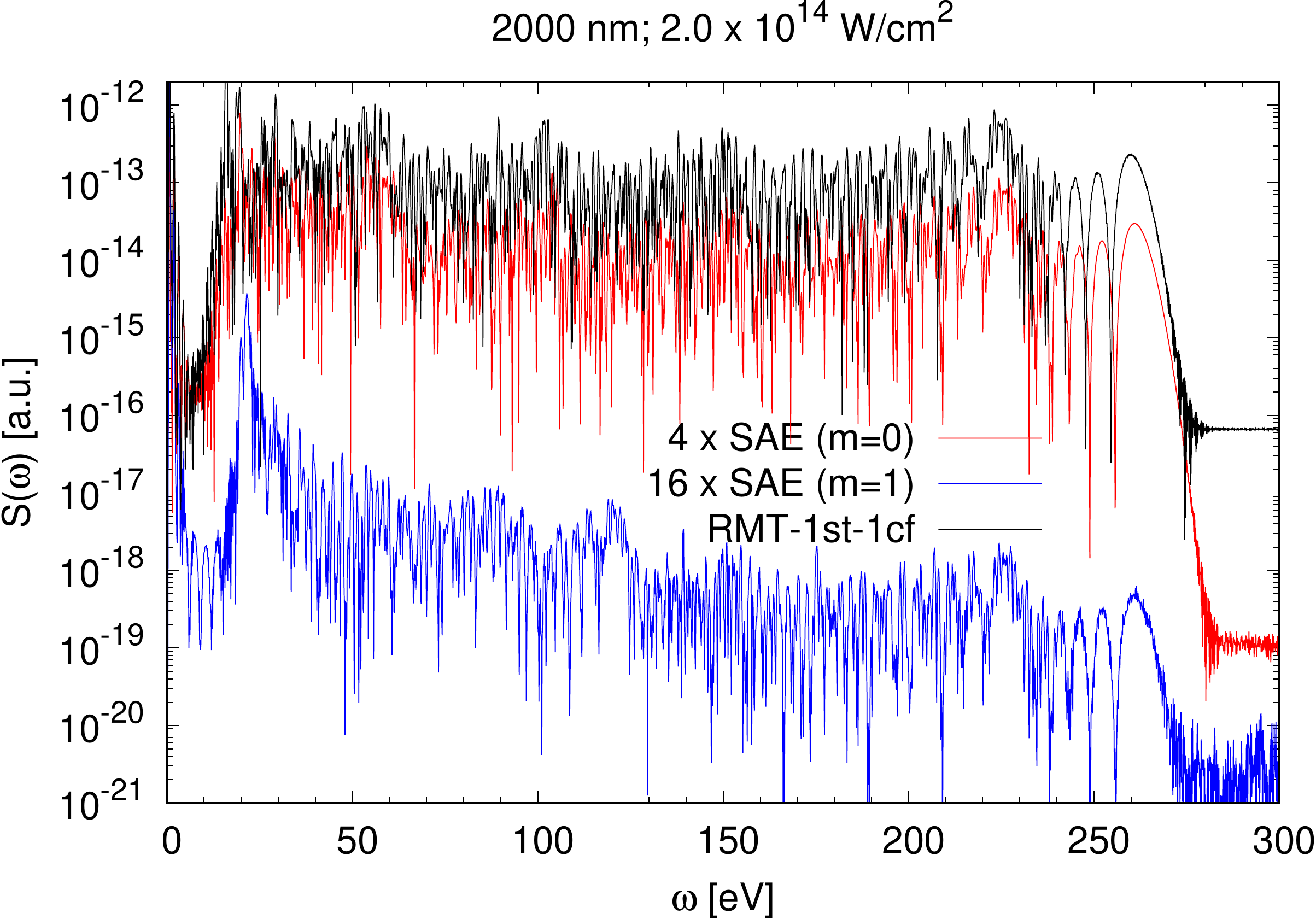}
    \caption{Harmonic spectrum of neon in a 2000~nm laser pulse with 
    a peak intensity of \intensity{2.0}{14}, as obtained in the 1st-1cf RMT and SAE models.  For the latter,
    the results for $m=0$ and $m=1$ are shown with the appropriate weighting factors.  See text for details.
    }
    \label{fig:2000nm}
\end{figure}

Having verified that the RMT calculations are both gauge-invariant and partial-wave converged, 
we can now investigate the results obtained using the SAE and
RMT methods, as well as the trends when varying the physical parameters of the calculations. 
\hbox{Figure~\ref{fig:2000nm}} shows a comparison of SAE and RMT results for a wavelength of 2000~nm and a peak
intensity of \intensity{2.0}{14}.  We notice good {\it qualitative\/} agreement between the two sets of predictions.  However,
there are substantial differences in the details, especially in light of the logarithmic scale.  As mentioned above,
the SAE results need to be weighted by the initial magnetic quantum number of the $2p$ orbital. However, the figure shows that
the contributions from $m=\pm 1$ are negligible compared to that from $m=0$.  We also see that both sets of 
results (RMT and SAE for $m=0$) show the plateau reaching up to about 260~eV, 
in excellent agreement with the prediction of 258~eV from
the \hbox{three-step} model.

\begin{figure}[!t]
    \centering
    \includegraphics[width=0.98\columnwidth]{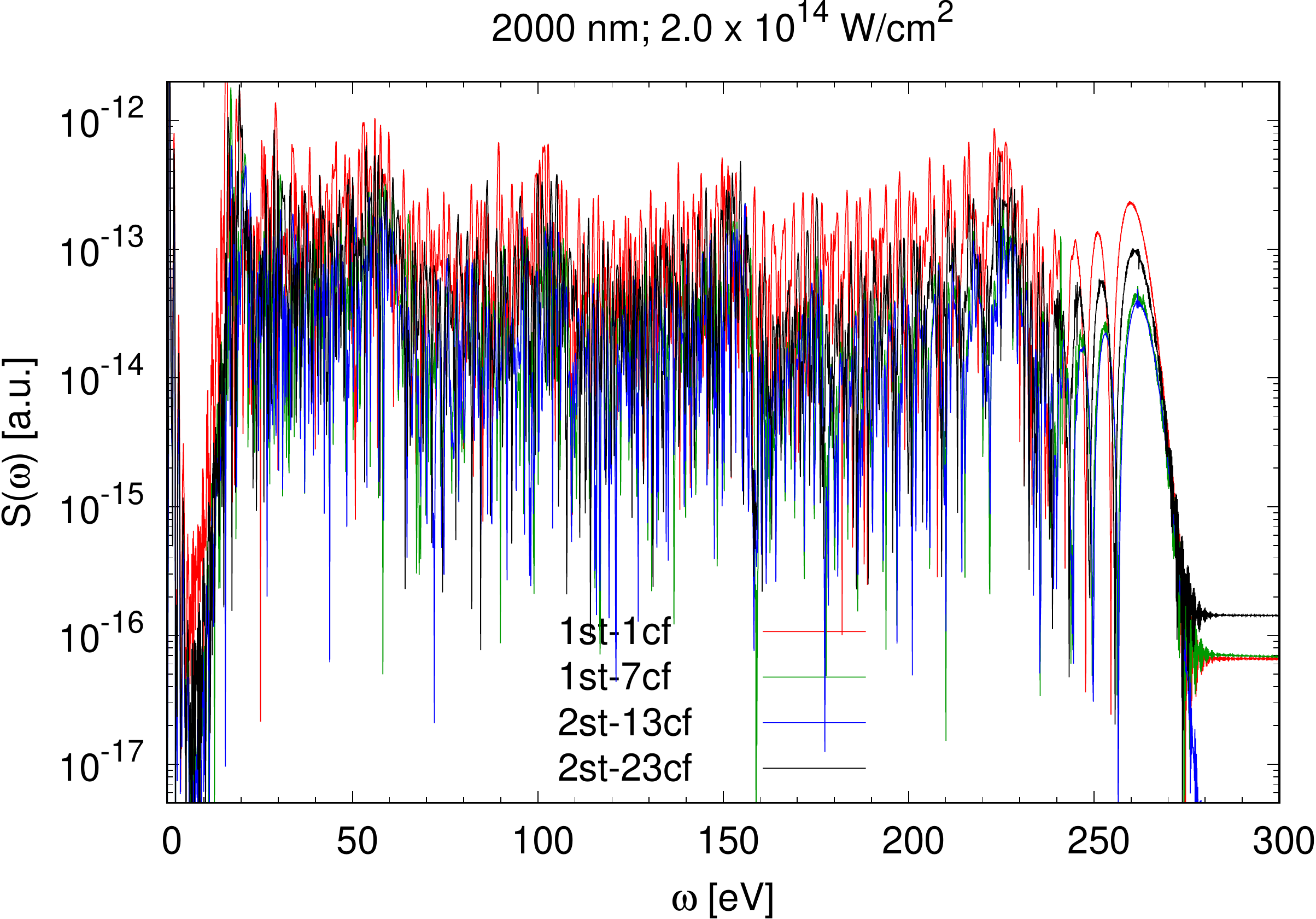}
    \caption{Harmonic spectrum of neon in a 2000~nm laser pulse with a
    peak intensity of \intensity{2.0}{14}. The results were obtained
    in the four RMT models described in the text. 
    }
    \label{fig:models}
\end{figure}

In light of the differences seen between the SAE results and those from the
simplest RMT model, i.e., 1st-1cf with the final ionic state described
by only its dominant configuration in the expansion, we now investigate the sensitivity of the results
on the number of coupled ionic states and their description. Our findings are
shown in Fig.~\ref{fig:models}. While the {qualitative\/} predictions are practically identical,
the {\it quantitative} results are not.

\begin{figure}[!t]
    \centering
    \includegraphics[width=0.98\columnwidth]{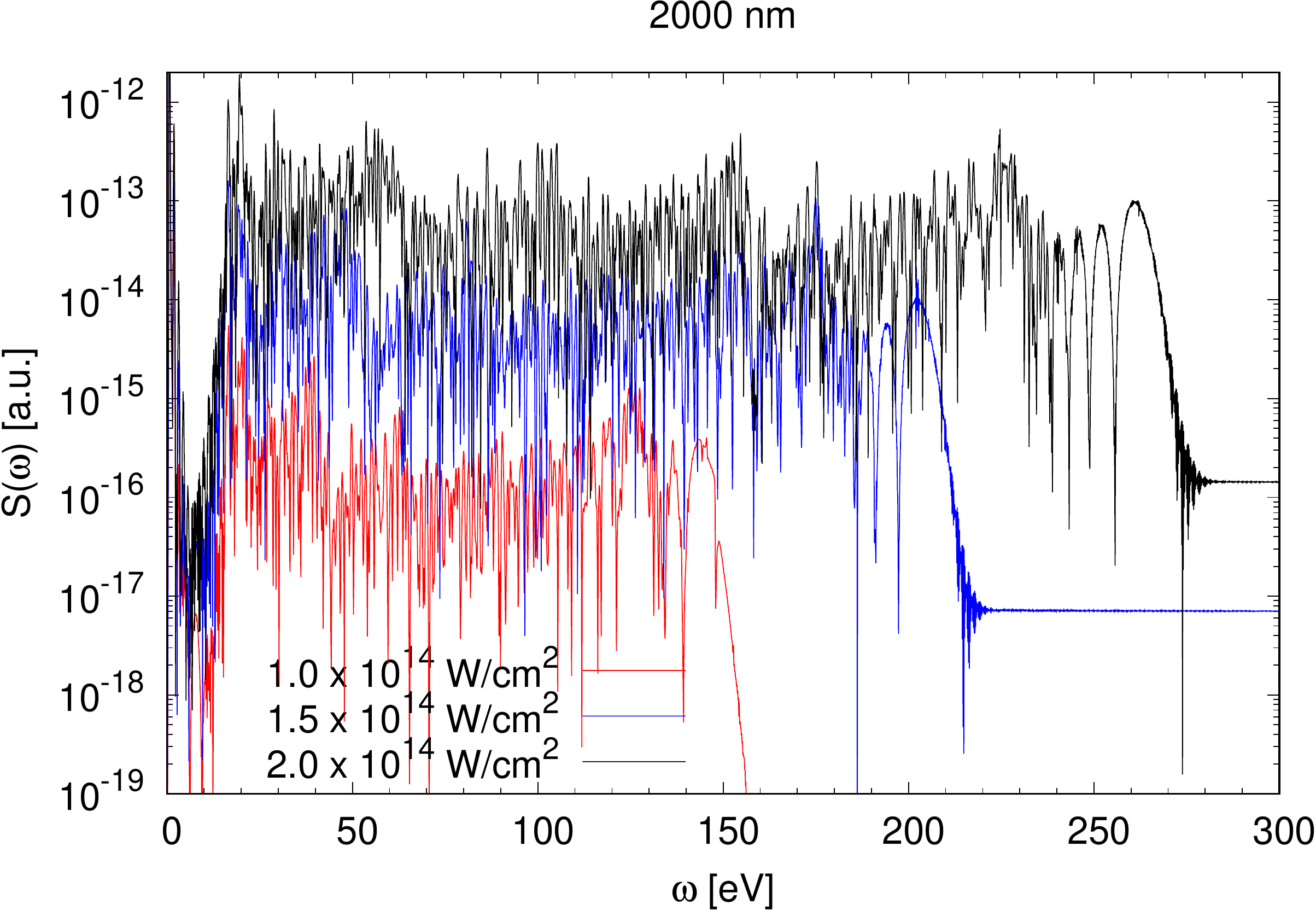}
    \caption{Harmonic spectra of neon in a 2000~nm laser pulse with peak
    intensities between \intensity{1.6}{14} and \intensity{2.0}{14}. The harmonic cutoffs 
    range from about 140~eV to 260~eV, following an approximately linear increase with intensity.
    }
    \label{fig:intensity}
\end{figure}

Since our principal goal is to test the RMT model, we now move to a discussion of 
the RMT results as a function of various experimentally controllable parameters.  
Figure~\ref{fig:intensity} shows that the harmonic cutoff 
changes approximately linearly with intensity, as expected from Eq.~(\ref{Up}). Furthermore, 
the calculated harmonic cutoffs agree well with those expected from Eq.~(\ref{cutoff_eq}). 
The largest disagreement is about 5~eV for the calculation with an intensity of \intensity{2.0}{14}.

\begin{figure}[!t]
    \centering
    \includegraphics[width=0.98\columnwidth]{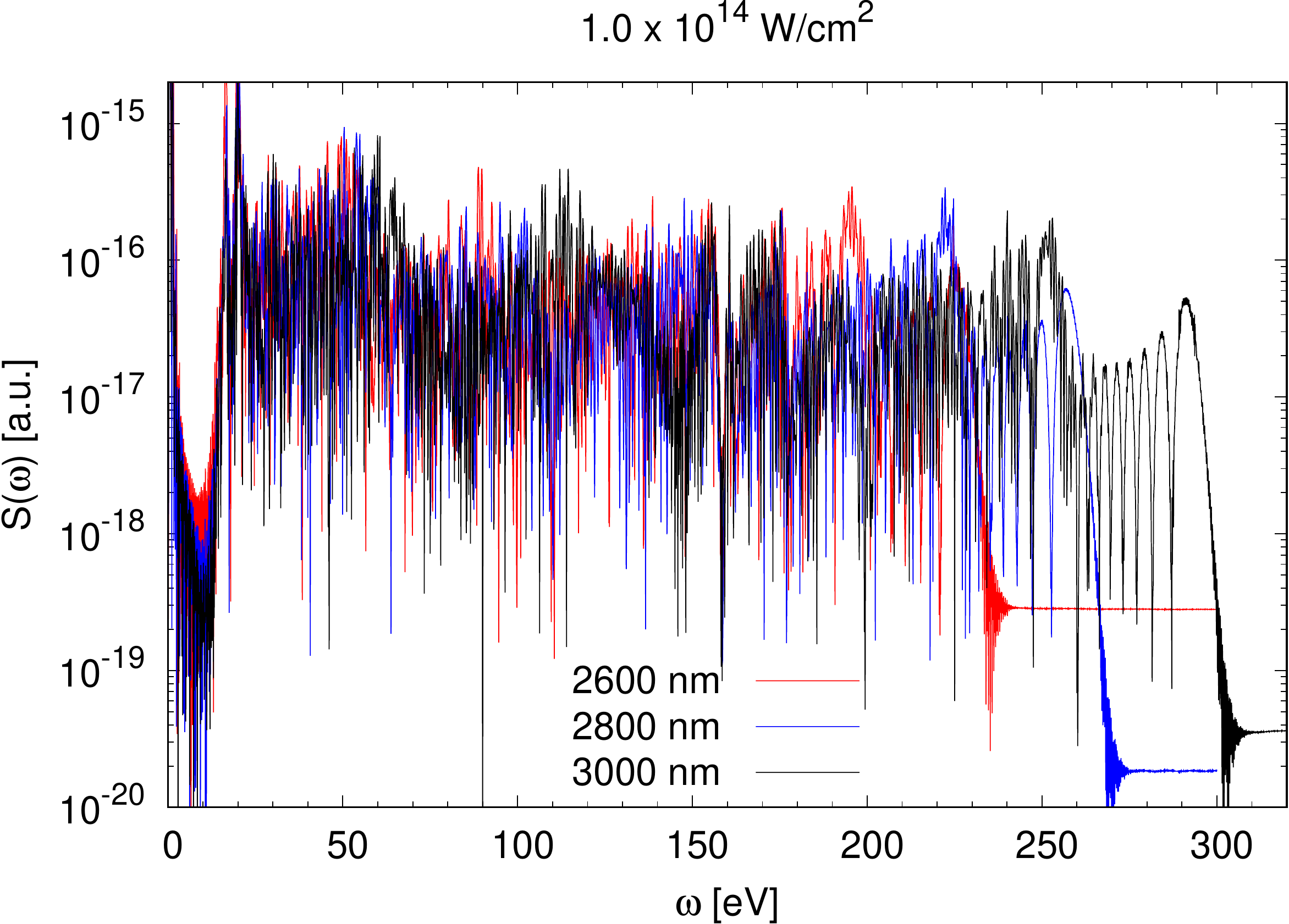}
    \caption{Harmonic spectra of neon in a \intensity{1.0}{14} laser pulse 
    with wavelengths ranging from 2600~nm to 3000~nm. As expected, the harmonic cutoffs vary from $\approx 220$~eV 
    to $\approx 290$~eV.
    }
    \label{fig:wavelength}
\end{figure}
Another expected trend based on Eq.~(\ref{Up}) is an approximately quadratic 
increase in the harmonic cutoff with increasing wavelength. The cutoffs in Fig.~\ref{fig:wavelength}, 
in general, agree with this expectation, with discrepancies of up to a few~eV. 
This is not unexpected, as Eq.~(\ref{Up}) does not account for the pulse envelope.

\begin{figure}[!tb]
    \centering
    \includegraphics[width=0.98\columnwidth]{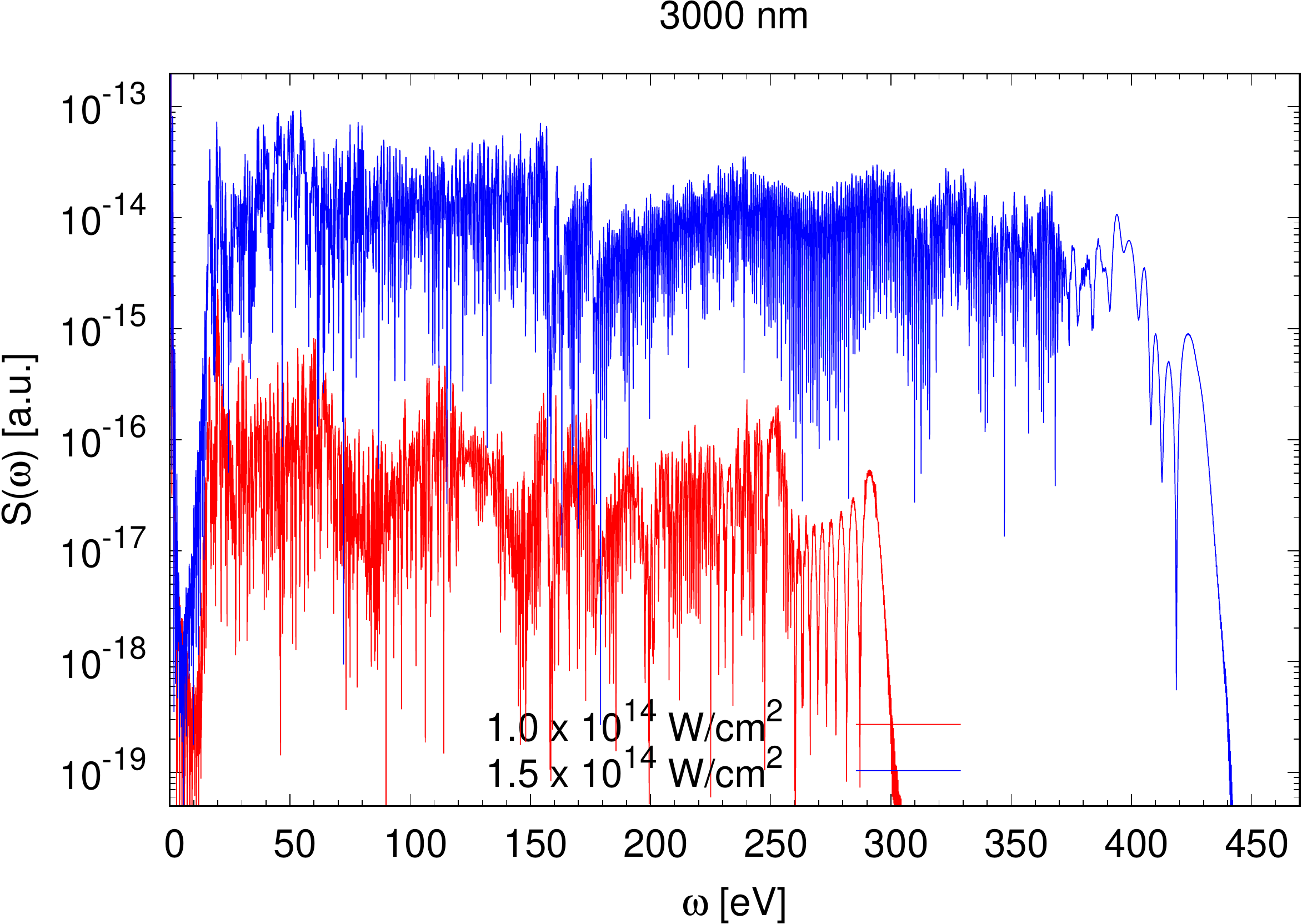}
    \caption{Harmonic spectra of neon from a 3000~nm pulse with peak intensities of 
    \intensity{1.0}{14} and \intensity{1.5}{14}. The latter produces a cutoff at $\approx 420$~eV.  
    }
    \label{fig:3000nm}
\end{figure}

Though the calculations shown so far exhibit the characteristics expected from 
Eqs.~(\ref{cutoff_eq}) and~(\ref{Up}), the cutoffs are not yet sufficiently high
to reach into the water window. 
As shown in Fig.~\ref{fig:3000nm}, increasing the wavelength to 3000~nm 
with a peak intensity of \intensity{1.5}{14} finally 
pushes the cutoff into this window.

\begin{table}[!t]
\begin{center}
\caption{Conversion efficiency in the various models. $C_t$ is obtained by integrating over the entire
spectrum until just above the cutoff, while the integration is limited to the plateau region for $C_p$.}
\vspace{0.2truemm}
\label{table:tab1}
\resizebox{.49\textwidth}{!}{
\begin{tabular}{c|c|c|c|c}
\hline
Model & $\lambda$  & Peak Intensity & $C_t$ & $C_p$\\ 
\hline
 SAE ($m=0$) & 2000 nm & \intensity{2.0}{14} & \num{3.50e-3} & \num{3.30e-3} \\
 SAE ($m=1$) & 2000 nm & \intensity{2.0}{14} & \num{3.00e-4} & \num{3.84e-9} \\
 1st-1cf & 2000 nm & \intensity{2.0}{14} & \num{4.20e-2} & \num{2.65e-2} \\
 1st-7cf & 2000 nm & \intensity{2.0}{14} & \num{2.27e-2} & \num{6.80e-3} \\
 2st-13cf & 2000 nm & \intensity{2.0}{14} & \num{2.07e-2} & \num{5.50e-3} \\
 2st-23cf & 2000 nm & \intensity{2.0}{14} & \num{3.02e-2} & \num{1.29e-2} \\
 2st-23cf & 2000 nm & \intensity{1.5}{14} & \num{1.86e-2} & \num{1.70e-3} \\
 2st-23cf & 2000 nm & \intensity{1.0}{14} & \num{1.77e-2} & \num{4.83e-5} \\
 2st-23cf & 2600 nm & \intensity{1.0}{14} & \num{6.20e-3} & \num{2.21e-5} \\
 2st-23cf & 2800 nm & \intensity{1.0}{14} & \num{4.30e-3} & \num{1.69e-5} \\
 2st-23cf & 3000 nm & \intensity{1.0}{14} & \num{3.30e-3} & \num{1.54e-5} \\
 2st-23cf & 3000 nm & \intensity{1.5}{14} & \num{6.00e-3} & \num{2.70e-3} \\
\hline
\end{tabular}
}
\end{center}
\end{table}

We finish our presentation with the conversion efficiencies obtained in the various models and choices of parameters.  These are listed in Table~\ref{table:tab1}.
As seen in the table, the predicted conversion efficiency depends significantly on the details of the model employed. In particular, the SAE results for 2000~nm at \intensity{2.0}{14} are much smaller than those obtained
with the more correlated RMT models.  Even within the latter models, however, the predicted
conversion efficiency $C_p$, obtained by integrating over the plateau region, 
varies between a minimum of 
$5.50\times 10^{-3}$ in \hbox{2st-13cf} and a maximum of $2.65\times 10^{-2}$ in
\hbox{1st-1cf}. The variability is less in $C_t$, which is obtained by integrating over
the entire spectrum. Apparently the models are less sensitive when it comes to 
predicting the dominant first peak near the driving frequency.  
This is the usual scenario that relatively large numbers are less affected by details
in the model. 

Next, it is important to investigate how the harmonic 
yield depends on the intensity and wavelength of the laser. As mentioned above, we expect the yield to decrease 
as the wavelength increases for fixed intensity. This expectation is somewhat indicated in Fig.~\ref{fig:wavelength} and clearly confirmed in Tab.~\ref{table:tab1}. Specifically, increasing the wavelength
from 2600~nm to 3000~nm at \intensity{2.0}{14} reduces the total conversion efficiency~$C_t$
by almost a factor of~2, which is consistent with an approximate
$\lambda^{-5}$ dependence.  However, the conversion efficiency~$C_p$
at 3000~nm is still about 70\% of the value at 2600~nm, thus suggesting
a smaller reduction.

\section{\label{sec:ch5conclusion}Conclusions and Outlook}

We have investigated the efficacy of the $R$-Matrix with Time-dependence (RMT) method to 
generate high-order harmonics in the water window using mid-IR laser light. Previous approaches 
relied on significant simplifications that neglect both quantum-mechanical and multi-electron 
effects on the predicted spectra. Since it is not clear how accurate the results of 
these simplified models are, the present project was designed to carry out a thorough test.

We successfully demonstrated that RMT is capable of generating harmonics in the water window. 
We found qualitative agreement with the trends predicted in classical models and SAE calculations regarding the overall frequency dependence of the spectral density, specifically the dominating
first few harmonics, the low-energy minimum, the plateau-like structure, and the cutoff frequency. 

However, we found a significant dependence of the absolute values of the spectral density, and
consequently the calculated conversion efficiency, on the 
details of the model employed. 
While employing windowing functions such as the Blackman window may improve the appearance of the spectrum, they significantly modify the shape of the harmonic plateau and therefore also the conversion efficiency, especially in few-cycle laser fields that are strongly ramped. Although it may seem unsatisfactory that we can only suggest that
the results from our largest model \hbox{(2st-23cf)} are the most reliable, we hope that
the detailed analysis provided in this paper will encourage further work and, 
in particular, motivate theorists to publish absolute numbers.

In the future, we plan to interface the RMT approach with the B-spline R-matrix (BSR) code 
of Zatsarinny~\cite{BSR3}, which allows the use of non\-orthogonal, term-dependent 
target orbitals. As demonstrated in numerous applications to time-independent treatment of
electron and photon collisions, we expect this to further improve the inner-region part of
RMT, especially for complex targets. As an ultimate test of the RMT method, 
we hope to compare our predictions with experimental data. This will likely require
substantial efforts by both the experimental and the theoretical communities. We hope that the 
work presented in this paper will stimulate such efforts.

\section*{\label{sec:acknowlegements}Acknowledgements}

This work was supported by the United States National Science Foundation 
under grants No.~PHY-1803844, OAC-1834740, PHY-2110023 (K.B.), and PHY-2012078 (N.D.). The calculations 
were carried out on Stampede2 and Frontera at the Texas Advanced Computing Center in Austin, 
with access provided by the XSEDE supercomputer allocation No.~PHY-090031 (Stampede2) and 
Frontera Pathways allocation PHY20028, and on Bridges-2 at the Pittsburgh Supercomputing 
Center through the Early User Program of the allocation No.~PHY-180023.  
We thank Anne Harth for useful discussions regarding the importance of calculating the conversion efficiency.
The RMT code is part of the UK-AMOR suite and can be accessed free of charge at~\cite{ukamorsite}.

\end{document}